\documentclass[prl,twocolumn,superscriptaddress,showpacs]{revtex4}
\usepackage{graphicx}

\newcommand*{\eg}{{e.g., }}

\newcommand*{\etal}{{\sl et al.}}
\newcommand*{\lta}{\lesssim}

\newcommand*{\cobe}{{COBE}}
\newcommand*{\dmr}{{DMR}}
\newcommand*{\boom}{{BOOMERANG}}
\newcommand*{\mxm}{{MAXIMA-1}}

\begin{document}
\bibliographystyle{apsrev}

\preprint{}

\title{Cosmology from MAXIMA-1, BOOMERANG \& COBE/DMR CMB Observations.}
\newcommand{\cita}{Canadian Institute for Theoretical Astrophysics,
  University of Toronto, Toronto M5S 3H8, Canada} 
\newcommand{\cfpa}{Center for Particle
  Astrophysics, University of California, Berkeley, CA 94720}
\newcommand{\lasapienza}{Dipartimento di Fisica, Universita' La
  Sapienza, Roma, Italy} 
\newcommand{\caltech}{California Institute of  Technology, Pasadena, CA 91125} 
\newcommand{\spacescience}{Space Sciences Laboratory, University of
  California, Berkeley, CA 94720} 
\newcommand{\ucsb}{Department of Physics, University of California, Santa
  Barbara, CA 93106} 
\newcommand{\ENEA}{ENEA Centro Ricerche di Frascati,
  Via E. Fermi 45, 00044 Frascati, Italy} 
\newcommand{\ucbphysics}{Department of Physics, University of California,
  Berkeley, CA 94720}
\newcommand{\ucbastro}{Department of Astronomy, University of California,
  Berkeley, CA 94720}


\author{A.H.~Jaffe}
\affiliation{\cfpa}\affiliation{\spacescience}
\affiliation{\ucbastro}
\author{P.A.R.~Ade}
\affiliation{Queen Mary and Westfield College, London, UK}
\author{A.~Balbi}
\affiliation{\cfpa}
\affiliation{Dipartimento di Fisica, Universit{\`a} Tor Vergata,
  Roma, Italy}
\affiliation{Division of Physics, Lawrence Berkeley National Laboratory, 
  Berkeley, CA 94720}
\author{J.J~Bock}
\affiliation{Jet Propulsion Laboratory, Pasadena, CA 91109}
\author{J.R.~Bond}
\affiliation{\cita}
\author{J.~Borrill}
\affiliation{\cfpa}
\affiliation{\spacescience}
\affiliation{\ucbastro}
\affiliation{National Energy Research Scientific Computing Center, LBNL,
  Berkeley, CA 94720}
\author{A.~Boscaleri}
\affiliation{IROE-CNR, Firenze, Italy}
\author{K.~Coble}
\affiliation{\ucsb}
\author{B.P.~Crill}
\affiliation{\caltech}
\author{P.~de~Bernardis}
\affiliation{\lasapienza}
\author{P.~Farese}
\affiliation{\ucsb}
\author{P.G.~Ferreira}
\affiliation{Astrophysics, University of Oxford, NAPL, Keble Road, OX2 6HT, UK}
\affiliation{CENTRA, Instituto Superior Tecnico, Lisbon, Portugal}
\author{K.~Ganga}
\affiliation{\caltech}
\affiliation{Physique Corpusculaire et Cosmologie, Coll{\`e}ge de France, 11
  Place Marcelin Berthelot, 75231 Paris Cedex 05, France}
\author{M.~Giacometti}
\affiliation{\lasapienza}
\author{S.~Hanany}
\affiliation{School of Physics and Astronomy, University of
  Minnesota/Twin Cities, Minneapolis, MN 55455}
\affiliation{\cfpa}
\author{E.~Hivon}
\affiliation{\cfpa}
\affiliation{\caltech}
\author{V.V.~Hristov}
\affiliation{\caltech}
\author{A.~Iacoangeli}
\affiliation{\lasapienza}
\author{A.E.~Lange}
\affiliation{\cfpa}
\affiliation{\caltech}
\author{A.T.~Lee}
\affiliation{\cfpa}
\affiliation{\spacescience}
\author{L.~Martinis}
\affiliation{\ENEA}
\author{S.~Masi}
\affiliation{\lasapienza}
\author{P.D.~Mauskopf}
\affiliation{University of Wales, Cardiff, UK, CF24 3YB}
\author{A.~Melchiorri}
\affiliation{\lasapienza}
\author{T.~Montroy}
\affiliation{\ucsb}
\author{C.B.~Netterfield}
\affiliation{Departments of Physics and Astronomy, University of
  Toronto, Toronto M5S 3H8, Canada} 
\author{S.~Oh}
\affiliation{\cfpa}
\affiliation{\ucbphysics}
\author{E.~Pascale}
\affiliation{IROE-CNR, Firenze, Italy}
\author{F.~Piacentini}
\affiliation{\lasapienza}
\author{D.~Pogosyan} 
\affiliation{\cita}
\author{S.~Prunet} 
\affiliation{\cita}
\author{B.~Rabii}
\affiliation{\cfpa}
\affiliation{\spacescience}
\affiliation{\ucbphysics}
\author{S.~Rao}
\affiliation{Istituto Nazionale di Geofisica, Roma, Italy}
\author{P.L.~Richards}
\affiliation{\cfpa}
\affiliation{\ucbphysics}
\author{G.~Romeo}
\affiliation{Istituto Nazionale di Geofisica, Roma, Italy}
\author{J.E.~Ruhl}
\affiliation{\cfpa}
\affiliation{\ucsb}
\author{F.~Scaramuzzi}
\affiliation{\ENEA}
\author{D.~Sforna}
\affiliation{\lasapienza}
\author{G.F.~Smoot}
\affiliation{\cfpa}
\affiliation{\spacescience}
\affiliation{\ucbphysics}
\affiliation{Division of Physics, Lawrence Berkeley National Laboratory, 
  Berkeley 94720, CA}
\author{R.~Stompor}
\affiliation{\cfpa}
\affiliation{\spacescience}
\affiliation{Copernicus Astronomical Center, Warszawa, Poland}
\author{C.D.~Winant}
\affiliation{\cfpa}
\affiliation{\spacescience}
\affiliation{\ucbphysics}
\author{J.H.P.~Wu}
\affiliation{\ucbastro}

\begin{abstract}
\vskip0.1in
  Recent results from \boom-98 and \mxm, taken together with \cobe-\dmr,
  provide consistent and high signal-to-noise measurements of the CMB
  power spectrum at spherical harmonic multipole bands over
  $2<\ell\lta800$.  Analysis of the combined data
  yields 68\% (95\%) confidence limits on the total density,
  $\Omega_{\rm {tot}}\simeq 1.11 \pm 0.07\ (^{+0.13}_{-0.12})$, the
  baryon density, $\Omega_b h^2\simeq 0.032^{+0.005}_{-0.004}\ 
  (^{+0.009}_{-0.008})$, and the scalar
  spectral tilt, $n_s\simeq1.01^{+0.09}_{-0.07}\ (^{+0.17}_{-0.14})$. 
  These data are consistent with inflationary initial conditions for
  structure formation. Taken together with other cosmological
  observations, they imply the existence of both non-baryonic dark
  matter and dark energy in the universe.
\end{abstract}


\maketitle

Measurements of the angular power spectrum, $\mathcal{C}_\ell$, of the
Cosmic Microwave Background (CMB) have long been expected to enable
precise determinations of cosmological parameters~\cite{cmbprediction}.
The CMB power spectrum depends on these parameters, as well as the
scenario for the generation and growth of density fluctuations in the
early universe.  Evidence for structure in the CMB of the character
predicted by adiabatic inflationary models has been mounting for the
past decade and was convincingly detected in
1999~\cite{toco+b97,melch,dodknox}.  The recent \boom-98
(B98)~\cite{BOOM1} and \mxm~\cite{MAXIMA1} CMB anisotropy data provide a
significant improvement in the determination of $\mathcal{C}_\ell$.
This letter jointly analyzes these two data sets, incorporating
\cobe-\dmr~\cite{DMR} and other cosmological information to obtain
further estimates of several cosmological parameters.

\begin{figure}[htb]\vspace{-0.1\columnwidth}
  \includegraphics[width=1.0\columnwidth]{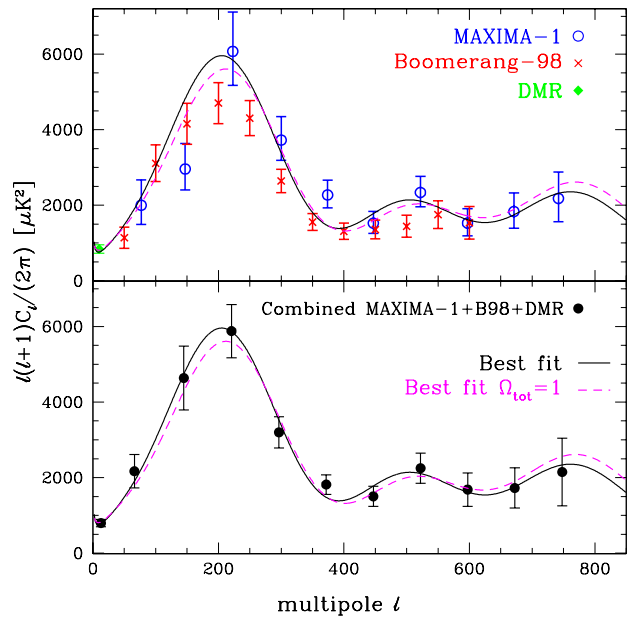}
  \caption{
    CMB power spectra, $\mathcal{C}_\ell =
    \ell(\ell+1)C_\ell/2\pi$.  Top: \mxm, B98\ and \cobe-\dmr.
    Bottom: maximum-likelihood fit to the power in bands for the three
    spectra, marginalized over beam and calibration uncertainty. In both
    panels the curves show the best fit model in the joint parameter
    estimation with weak priors and the best fit with $\Omega_{\rm
      {tot}}=1$. These models have \{$\Omega_{\rm {tot}}$,
    $\Omega_\Lambda$, $\Omega_b h^2$, $\Omega_c
    h^2$,$n_s$,$\tau_C$\}$=$\{1.2, 0.5, 0.03, 0.12, 0.95, 0\}, \{1, 0.7,
    0.03, 0.17, 0.975, 0\}. They remain the best fits when the large
    scale structure prior~\cite{LSS} is added, and when the SN
    prior~\cite{SNIa} is added the $\Omega_{\rm {tot}}=1$ model becomes
    the best fit in both cases. }
  \label{fig:powerspectrum}  
\end{figure}

\mxm\ and B98 have produced independent power spectra from patches of
sky roughly $90^\circ$ apart, on opposite sides of the galactic plane
(\cite{BOOM1,MAXIMA1} and references therein).  These data provide the
first narrow-band detections of the power spectrum from
$400\lta\ell\lta800$, where further acoustic peaks are expected in
adiabatic models.  Each spectrum shows a well-defined peak at multipole
$\ell\sim200$, followed by a relatively flat region extending to the
highest multipoles reported~(Figure~\ref{fig:powerspectrum}, top).
These results have been interpreted as supporting the inflationary
theory of structure formation with adiabatic initial conditions, and
allow the first precise CMB measurements of other parameters such as the
baryon density \cite{BOOM2,MAXIMA2,otherboominterp}.

\paragraph*{Comparison \& Calibration.}The B98\ and \mxm\ angular
power spectra~\cite{bandpowpost} are shown in the top panel of
Fig.~\ref{fig:powerspectrum}, along with some best-fit models. The B98
(\mxm) data cover $\ell=25$--$625$ (36--785) with a resolution
$\delta\ell=50$ (75).  The \dmr\ data provide information at low $\ell$.
Each data set has a calibration uncertainty (20\% for B98, 8\% for \mxm,
1$\sigma$ in $\mathcal{C}_\ell$) and a beam uncertainty (10\% for B98,
5\% for \mxm) that are not included in the errors plotted in
Fig.~\ref{fig:powerspectrum}. We take these uncertainties into account
by allowing the overall amplitude (calibration) as well as an
$\ell$-dependent amplitude (beam) to vary for each spectrum, weighted by
the distribution of errors around the nominal beam and calibration
values.

We have combined the two data sets in the bottom panel of
Fig.~\ref{fig:powerspectrum}.  We approximate the likelihood in the
manner of~\cite{bjk00,dodknox}, treating the individual bandpowers as
the parameters to be determined.  The combined power spectrum requires
an increase in the calibration by a factor $1.14\pm 0.10$ for B98\ and a
decrease by $0.98\pm 0.08$ for \mxm, along with a beam rescaling of
$1.07\pm 0.09$ for B98\ and $0.99\pm 0.05$ for \mxm.

We define a goodness-of-fit parameter, $\chi^2\equiv-2\ln{\mathcal{L}}$,
in terms of the likelihood function, $\mathcal{L}$, which reduces to the
usual $\chi^2$ for a Gaussian.  The data shown in the top panel of
Fig.~\ref{fig:powerspectrum} have $\chi^2/{\rm DOF}=11.6/8$ with respect
to the combined spectrum of the bottom panel (22 data points, 10
bandpowers and 2 beam rescalings and 2 calibration parameters give 8
degrees of freedom [DOF]) when \dmr\ is not included, and a $\chi^2/{\rm
  DOF}=34/29$ with \dmr.  This indicates that the results are each
consistent with a single underlying power spectrum.

\paragraph*{Cosmological Parameters.}We consider a subset of
parameters describing a Friedmann-Robertson-Walker Universe with
adiabatic initial conditions.  The present-day density $\rho_i$ of a
component $i$ is $\Omega_i = 8\pi G\rho_i/(3H_0^2)$ where
$H_0=100h\;$km/s/Mpc is the Hubble constant; thus $\Omega_i h^2$ is a
physical density, independent of $H_0$.  We consider the density of
baryons, $\Omega_b$; the density of cold dark matter (CDM), $\Omega_c$;
the total density of matter, $\Omega_m=\Omega_c+\Omega_b$; the effective
density of a cosmological constant, $\Omega_\Lambda$; and finally the
total density $\Omega_{\rm {tot}}=\Omega_m+\Omega_\Lambda$, ignoring
other possible components.  The initial spectrum of density
perturbations is described by an amplitude factor, $\mathcal{C}_{10}$,
multiplying the CMB spectrum, and the spectral tilt of scalar (density)
perturbations, $n_s$ (defined so the initial three-dimensional
perturbation spectrum is $P(k)\propto k^{n_s}$).  We also consider the
optical depth to the epoch of reionization, $\tau_C$.  Our parameter
space is thus $\{\Omega_{\rm {tot}}$, $\Omega_\Lambda$, $\Omega_b h^2$,
$\Omega_c h^2$, $n_s$, $\tau_C$, $\ln\mathcal{C}_{10}\}$
\cite{otherparams}.

Because many of these parameters affect the spectrum in highly
correlated (in some cases almost degenerate) ways, {\em limits on any
  one parameter are necessarily a function of the constraints, implicit
  and explicit, that one assumes for the other parameters}. Thus, all
such limits must be understood in the context of the specific question
that one asks of the data.

Inflation  predicts $\Omega_{\rm {tot}} = 1$. Fig.~\ref{fig:powerspectrum}
shows that the data are consistent with this prediction: there exist
$\Omega_{\rm {tot}}=1$ models that fit the data well with quite
reasonable choices for all of the other parameters.  Simultaneously
maximizing the likelihood for all parameters at once gives $\Omega_{\rm
  {tot}} = 1.2$.  As is clear from the Figure, the global best-fit model
and the best-fit flat model both fit the data with $\chi^2/{\rm
  DOF}\lta1$.  Extracting quantitative conclusions from a simple
$\chi^2$ analysis is complicated by the fact that parameters are
correlated with one another and are bounded by other cosmological
observations.  This issue will be explored elsewhere.

In this paper, we constrain parameters individually by marginalizing
(integrating the posterior distribution) over all other parameters
(including those describing uncertainty in the beam and calibration). We
apply the techniques used on B97~\cite{melch}, B98~\cite{BOOM2} and
\mxm~\cite{MAXIMA2} to the combined data.  We calculate the likelihood
of a suite of models given the \dmr, \mxm, and B98\ results, taking into
account the correlations within each data set, as well as the
non-Gaussianity of the likelihood functions~\cite{bjk98,bjk00}.  We use
the median, 16\% and 84\% integrals as central and $\pm1\sigma$ error
estimates respectively.  $95\%$ limits are approximately twice the
quoted errors.

The most likely value of each parameter calculated in this way will, in
general, be different from that found by multi-dimensional maximization.
Marginalization has the advantage of taking into account the likelihood
structure over the entire space, weighted by the likelihood value and
the volume of the parameter space.  It has the disadvantage of allowing
models whose validity we may doubt to influence the final result.
Hence, it is crucial to restrict the range that each parameter may take
on through the prudent use of ``prior'' constraints, and to test how the
results change under a change of priors.

The finite range of each parameter that is included in our model
database acts as a uniform (tophat) prior probability
density~\cite{paramlims}.  We further restrict the analysis to the
cosmologically interesting regime of $0.45 < h < 0.90$, age
$t_0>10$~Gyr, and $\Omega_m>0.1$; we refer to this combination as our
``weak prior.'' Without such restrictions the data sets allow
pathological low-sound-speed models with strong positive curvature
($\Omega_k<0$), very high baryon density, and very young ages.
Parameter degeneracies allow such models to contribute to or dominate
the likelihood~\cite{BOOM2,degeneracies}.  The results presented here
were checked using the methods of~\cite{MAXIMA2,melch} and found to be
in good agreement~\cite{apj2diffs}.

Results are shown in Table~\ref{tab}.  The combined data yield
parameters consistent with those derived individually: the curvature is
close to flat, the spectral index is close to unity, and the baryon
density is about 2$\sigma$ above (with respect to our CMB confidence
limits) the favored Big-Bang nucleosynthesis (BBN) value of $\Omega_b
h^2=0.019\pm0.002$~\cite{bbn}.  This high value
suppresses the second acoustic peak relative to standard CDM models.

\begingroup\squeezetable
\begin{table}[bt]
\renewcommand*{\arraystretch}{1.5}
\caption{
  Parameter estimates from the two
  data sets~\cite{apj2diffs}, and the combined data, using the weak 
  prior ($0.45 < h < 0.90$, $t_0>10$~Gyr, $\Omega_m>0.1$).
  Below the line,
  we restrict the parameter space 
  to $\Omega_{\rm {tot}}=1$ and add other cosmological information.
  Central values and $1\sigma$ limits 
  are found from the 50\%, 
  16\% and 84\% integrals of the marginalized likelihood. 
  $\tau_C$ and $\Omega_\Lambda$ are not constrained by the data.
}\label{tab}
\begin{ruledtabular}
\begin{tabular}{lrrrrrrrr}
& \multicolumn{1}{c}{$\Omega_{\rm {tot}}$}
& \multicolumn{1}{c}{$\Omega_bh^2$}
& \multicolumn{1}{c}{$n_s$}
& \multicolumn{1}{c}{$\Omega_c h^2$}
\\
\colrule
B98+\dmr
&$1.15^{+0.10}_{-0.09}$  
&$0.036^{+0.006}_{-0.005}$
&$1.04^{+0.10}_{-0.09}$   
&$0.24^{+0.08}_{-0.09}$ 
\\
\mxm+\dmr
&$1.01^{+0.09}_{-0.09}$
&$0.031^{+0.007}_{-0.006}$
&$1.06^{+0.10}_{-0.09}$
& $0.18^{+0.07}_{-0.06}$
\\
B98+\mxm+\dmr
&$1.11^{+0.07}_{-0.07}$
& $0.032^{+0.005}_{-0.005}$
& $1.01^{+0.09}_{-0.08}$
& $0.14^{+0.06}_{-0.05}$
\\
\hline
\multicolumn{1}{r}{+($\Omega_{\rm {tot}}=1)$}
& 1
& $0.030^{+0.004}_{-0.004}$
& $0.99^{+0.07}_{-0.06}$
& $0.19^{+0.07}_{-0.06}$
\\
\multicolumn{1}{r}{CMB+LSS}
& $1.11^{+0.05}_{-0.05}$
& $0.032^{+0.004}_{-0.004}$
& $1.00^{+0.09}_{-0.06}$
& $0.13^{+0.02}_{-0.01}$
\\
\multicolumn{1}{r}{CMB+SNIa}
& $1.09^{+0.06}_{-0.05}$
& $0.032^{+0.005}_{-0.005}$
& $1.00^{+0.09}_{-0.08}$
& $0.10^{+0.04}_{-0.04}$
\\
\multicolumn{1}{r}{CMB+LSS+SNIa}
& $1.06^{+0.04}_{-0.04}$
& $0.033^{+0.005}_{-0.004}$
& $1.03^{+0.09}_{-0.07}$
& $0.14^{+0.03}_{-0.02}$
\end{tabular}
\end{ruledtabular}
\end{table}
\endgroup

\begin{figure}[htb]\vspace{-0.1\columnwidth}
    \includegraphics[width=1.05\columnwidth]{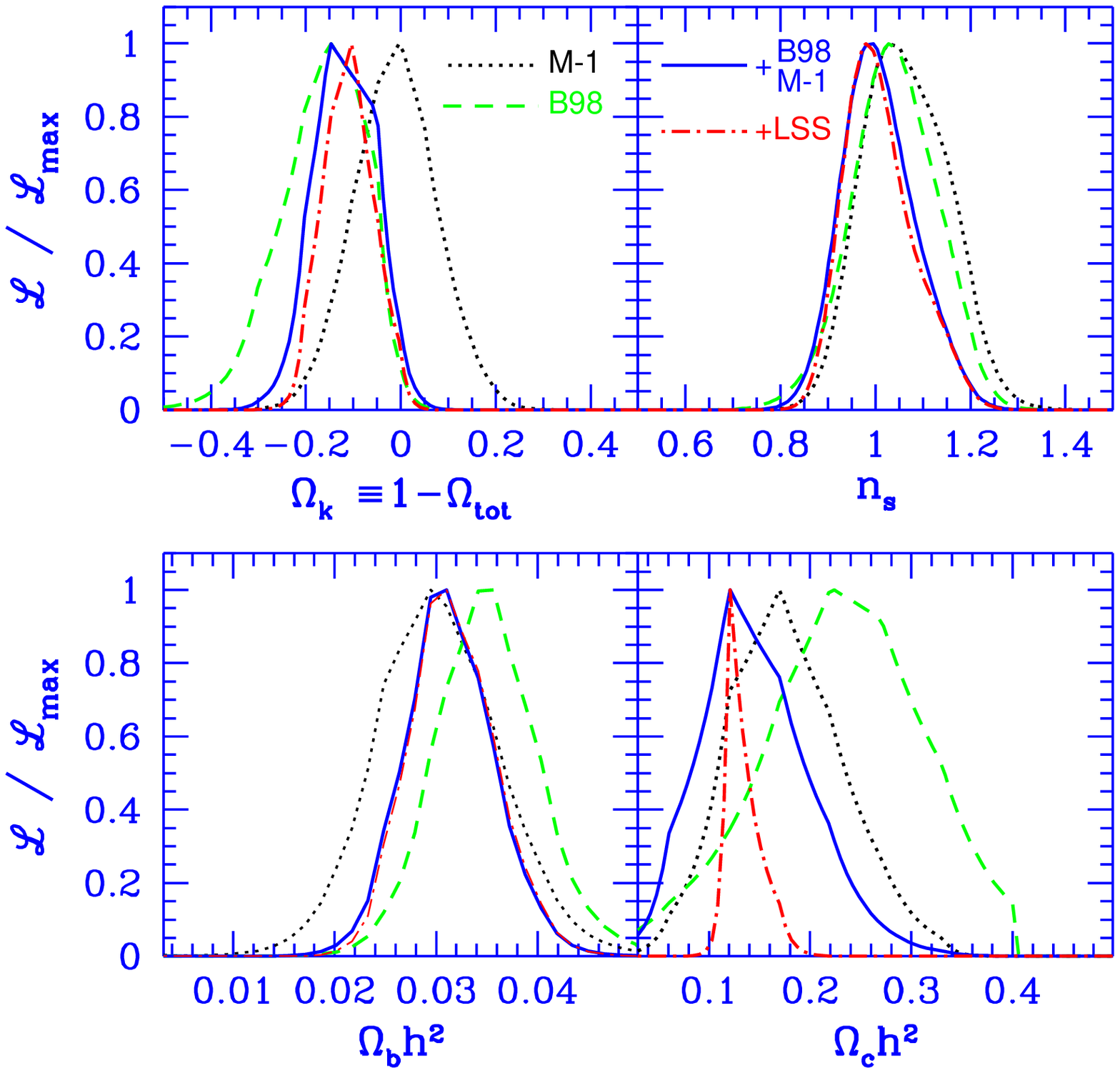}\vskip-0.8cm
    \includegraphics[height=0.5\columnwidth,angle=90]
    {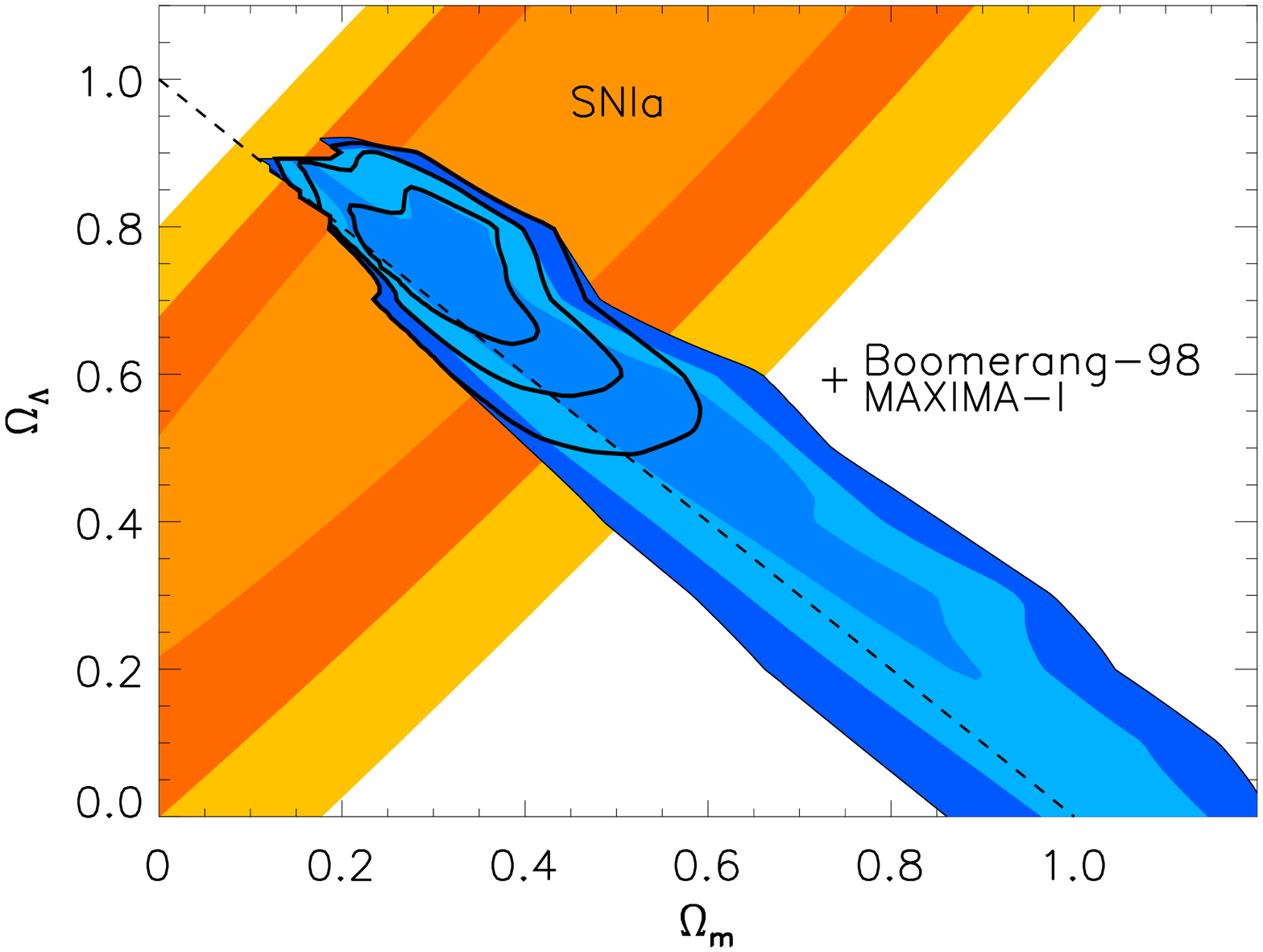}\hskip-0.95cm\includegraphics[
    height=0.5\columnwidth,angle=90]{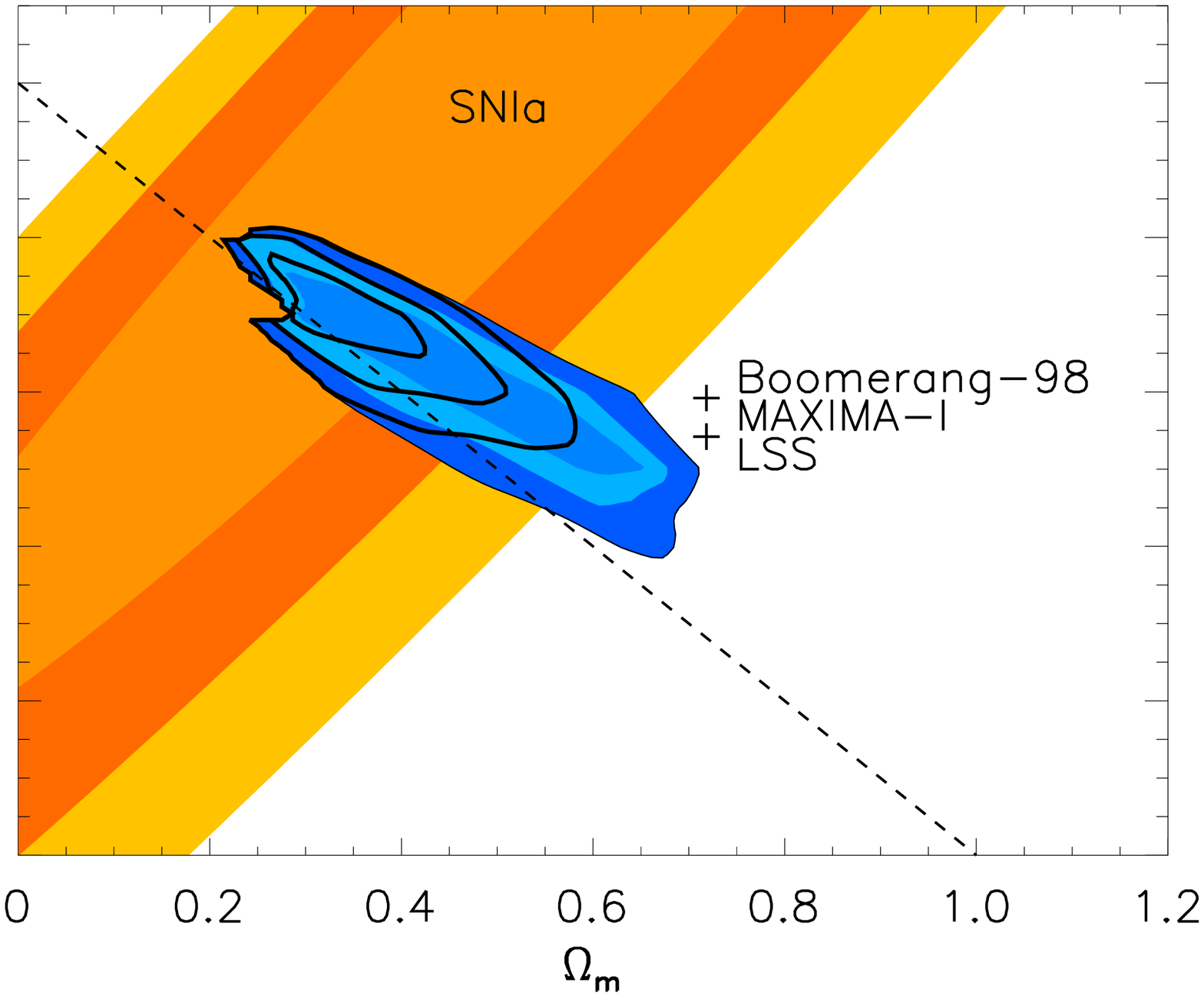}
    \vskip-0.5cm
    \caption{
      Likelihood functions calculated using the weak prior.  Top:
      likelihoods from \dmr+B98, \dmr+\mxm (M-1) \cite{apj2diffs},
      \dmr+\mxm+B98 and \dmr+\mxm+B98+LSS (LSS is the large-scale
      structure prior~\cite{LSS}).  Bottom: the likelihood in
      ($\Omega_m, \Omega_\Lambda$).  Shaded contours nearly parallel to
      $\Omega_m+\Omega_\Lambda=1$ are one-, two-, and three-sigma
      limits, defined as the equivalent likelihood ratio for a
      two-dimensional Gaussian distribution, from \dmr+B98+\mxm\ with
      weak priors (left) and \dmr+B98+\mxm+LSS (right).  Contours
      labeled ``SNIa'' are from high-redshift supernovae
      observations~\cite{SNIa}, and the final heavy set of contours are
      constraints from the product of the two distributions.  }
\label{fig:like}
\end{figure}

In Figure~\ref{fig:like} we plot likelihoods for several parameters
using the two data sets independently~\cite{apj2diffs}, and for the joint
analysis.  We also plot likelihood contours, showing how the CMB
primarily restricts $\Omega_{\rm {tot}}=\Omega_m+\Omega_\Lambda$.
Projecting this figure onto either axis gives the likelihood for the
quantities separately. This indicates the danger of individual parameter
estimates in the presence of strong correlations: for example, neither
$\Omega_m$ nor $\Omega_\Lambda$ are well-determined individually by the
CMB data alone.

We have also investigated the effects of applying various prior
probabilities~\cite{BOOM2}. Overall, the results from the combined data
are somewhat less dependent on the priors than either individually. We
find that any prior restriction that does not seriously contradict the
locus of good-fitting models has very little effect on the parameter
estimates.  Incorporating previous CMB data~\cite{bjk00,dodknox,BOOM2}
has very little effect.
Restricting the Hubble constant to
$h=0.71\pm0.08$~\cite{hubbleconst} 
does push the curvature closer to flat:
$\Omega_{\rm {tot}}=1.05^{+0.04}_{-0.04}$. A strong BBN prior ($\Omega_b
h^2 = 0.019 \pm 0.002$ ~\cite{bbn}), disfavored by the current CMB data, pushes
other quantities to compensate, such as $n_s=0.89\pm0.06$.

We determine parameters which are functions of those used to explicitly
define our parameter space by calculating their means and variances over
the full probability distribution.  We find $h=0.57\pm0.11$
and age $t_0=14.6\pm2.0$~Gyr.  Restricting to
$\Omega_{\rm {tot}} = 1$ gives $h=0.75\pm0.10$, 
$t_0=11.8\pm0.8$~Gyr.  To illustrate the effect of the priors
alone, we give the following limits without including CMB data:
$h=0.63\pm0.11$, 
$t_0=12.3\pm1.9$~Gyr.
 
The combined data improve the constraint on the matter density and
cosmological constant, but we caution that these results are influenced
by the weak prior alone (e.g., $\Omega_c h^2 = 0.18^{+0.11}_{-0.10}$
without CMB data).  However, prior information from other cosmological
data tightens these limits considerably. We consider the constraints on
the power spectrum from observations of large-scale-structure
(LSS)~\cite{LSS} and from observations of distant Supernova Ia
(SNIa)~\cite{SNIa} as in the bottom of Fig.~\ref{fig:like}. CMB+LSS
gives $(\Omega_m, \Omega_\Lambda)=(0.49\pm0.13, 0.63^{+0.08}_{-0.09})$;
(note that $\Omega_m$ is a function of the explicit database parameters,
and thus its confidence limits are calculated as discussed above)
CMB+SNIa gives $(0.35\pm0.07, 0.75^{+0.06}_{-0.07})$; combining all
three gives $(0.37\pm0.07, 0.71\pm{0.05}$ for CMB+LSS+SNIa.  We
note that the combination of CMB+LSS is about as restrictive as---and
compatible with---the combination of CMB+SNIa.

\paragraph*{Conclusions.}The \mxm\ and B98 data are consistent with
one another over the range of overlapping coverage in $\ell$.  The
consistency between the two data sets, obtained by different experiments
using different observation strategies on different parts of the sky,
eliminates many sources of systematic error as cause for concern.

These data, together with those of COBE-DMR, support the chief
predictions of the inflation paradigm, that the geometry of the universe
is flat, and that the initial density perturbations are scale-invariant,
and with the corollary that the density of mass-energy in the universe
is dominated by a form other than ordinary matter.  Simple models of
topological defects driving structure formation are difficult to
reconcile with the CMB data~\cite{strings}. These conclusions are
considerably strengthened by the inclusions of other cosmological data,
such as measurements of the Hubble constant, the amplitude and shape of
the matter power spectrum, and the accelerating expansion rate indicated
by observations of distant Supernovae.

Marginalization of the combined data over all other parameters yields a
value for the physical density of baryons of $\Omega_b
h^2=0.032\pm0.005$, ($0.030\pm0.004$ if $\Omega_{\rm {tot}}=1$).  These
results are each more than $2\sigma$ higher than the values determined
from the relative abundance of light elements and the theory of
BBN~\cite{bbn}.

The data analyzed here present only a suggestion of the expected second
acoustic peak in the CMB power spectrum.  A detailed mapping of that
$\ell$ region with higher signal-to-noise may yield more convincing
detections of acoustic oscillations, providing yet more
evidence for the adiabatic inflationary paradigm and measurements of the
cosmological parameters.  Already, the combined data begin to limit the
CDM density, driven by the B98\ constraints on the first peak and the
\mxm\ constraints over $650\lta\ell\lta800$.  Both teams are analyzing
additional data which may significantly reduce the errors in the region
of the power spectrum where further peaks are expected.

The \mxm\ and B98\ teams 
acknowledge support from the NSF through the Center for Particle
Astrophysics at UC Berkeley, the NSF Office of Polar Programs, the NSF
KDI program, from NASA, and from the DoE through NERSC in the USA; 
PPARC in the UK; and Programma Nazionale Ricerche in
Antartide, Agenzia Spaziale Italiana and University of Rome La Sapienza
in Italy. We thank the High-Z SN team and Saurabh Jha and Peter
Garnavich in particular for supplying SNIa likelihoods and for useful
conversations.


\vskip0.1in

\begin{thebibliography}{99}
  
\bibitem{cmbprediction} For example, G.\ Jungman \etal, \prl {\bf 76}, 1007
  (1996); G.\ Jungman \etal, \prd {\bf 54}, 1332 (1996); J.R.\ Bond, G.\ 
  Efstathiou \& M.\ Tegmark, Mon.\ Not.\ R.\ Astron.\ Soc.\ {\bf 291}, L33
  (1997) and references therein.

\bibitem{toco+b97} A.D.\ Miller \etal, \apj {\bf 524}, L1 (1999)
  astro-ph/9906421; P.\ Mauskopf \etal, \apj {\bf 536}, L63-L66 (2000),
  astro-ph/9911444.

\bibitem{melch} A.\ Melchiorri \etal, \apj {\bf 536}, (2000), L59-L62
  astro-ph/9911445.
  
\bibitem{dodknox} S.\ Dodelson \& L.\ Knox, \prl {\bf 84}, 3523 (2000),
  astro-ph/9909454.
  
\bibitem{BOOM1} P.\ de Bernardis \etal, \nat {\bf 404}, 995 (2000),
  astro-ph/0004404.  
  

\bibitem{MAXIMA1} S.\ Hanany \etal, \apj {\bf 545}, L5 (2000),
  astro-ph/0005123.

\bibitem{DMR} C.\ Bennett \etal, \apj {\bf 464}, L1 (1996).

\bibitem{BOOM2} A.E.\ Lange \etal, \prd submitted (2000), astro-ph/0005004
\bibitem{MAXIMA2} A.\ Balbi \etal, \apj {\bf 545}, L1 (2000),
  astro-ph/0005124. 

\bibitem{otherboominterp} M.\ White, D.\ Scott \& E.\ Pierpaoli, \apj Lett.,
  submitted (2000), astro-ph/0004385; M.\ Tegmark \& M.\ Zaldarriaga,
  \prl {\bf 85}, 2240 (2000), astro-ph/0004393. 
  
\bibitem{bjk00} J.R.\ Bond, A.H.\ Jaffe  \& L.\ Knox,
  \apj {\bf 533}, 19 (2000); astro-ph/9808264.
  
\bibitem{bandpowpost} The individual bandpowers will be available at
  http://cfpa.berkeley.edu/maxima,
  http://oberon.roma1.infn.it/boomerang, and
  http://www.physics.ucsb.edu/$\sim$boomerang/.

  
\bibitem{otherparams} We ignore several other parameters not constrained
  by these data.  The quintessence
  equation-of-state~\protect{\cite{darkenergy}} is largely degenerate
  with $\Omega_\Lambda$ and $\Omega_m$; the neutrino density has only a
  tiny effect on $C_\ell$ in the regime probed here; the gravity-wave
  amplitude awaits polarization measurements before it can be
  disentangled from the scalar amplitude.
  
\bibitem{darkenergy} For example, M.S.\ Turner in {\sl Proceedings of
    Physics in Collision}, edited by M.\ Campbell \& T.M.\ Wells (World
  Scientific, NJ, 2000), astro-ph/9912211; P.\ Brax, J.\ Martin \& A.\ 
  Riazuelo, \prd {\bf 62}, 103505 (2000), astro-ph/0005428.

\bibitem{bjk98}  J.R.\ Bond, A.H.\ Jaffe  \& L.\ Knox, \prd
 {\bf 57}, 2117 (1998), astro-ph/9708203.
  
\bibitem{paramlims} The amplitude $\mathcal{C}_{10}$ is a continuous
  variable.  The rest are discretized over: $0.1
  \leq\Omega_{\rm {tot}} \leq 1.5$; $0.0031\leq \Omega_{b}h^2 \leq
  0.2$; $0.03 \leq \Omega_c h^2 \leq 0.8$; $0 \leq \Omega_\Lambda \leq
  1.1$; $0 \leq \tau_C \leq 0.5$; $0.5 \leq n_s \leq 1.5$. A full
  discussion of the effect of limits and other priors on parameter
  determinations is in~\protect{\cite{BOOM2}}.

\bibitem{degeneracies} G.\ Efstathiou \& J.R.\ Bond,
  Mon.\ Not.\ R.\ Astron.\ Soc.\ {\bf 304}, 75 (1999).

\bibitem{apj2diffs} The small differences between the \mxm\ results here
  and in~\protect{\cite{MAXIMA2}} can be ascribed to differences in the
  parameter ranges and database gridding; restricting the parameter
  limits gives more consistent results and provides a check on the
  methods.
  
\bibitem{bbn} K.A.\ Olive, G.\ Steigman \& T.P.\ Walker, Phys.\ Rep.\ 
  {\bf 333-334}, 389 (2000), astro-ph/9905320; S.\ Burles, K.\ Nollett, J.\ 
  Truran \& M.S.\ Turner, \prl {\bf 82}, 4176 (1999), astro-ph/9901157;  D.\ 
  Tytler, J.M.\ O'Meara, N.\ Suzuki \& D.\ Lubin, Physica Scripta
  submitted, astro-ph/0001318 (2000); S.\ Burles, K.\ Nollett \& M.S.\ 
  Turner, \prd (2001), astro-ph/0008495, gives a slightly higher value
  of $\Omega_b h^2 = 0.020 \pm 0.002$.
  
\bibitem{hubbleconst} W.L.\ Freedman, Phys.\ Rep. {\bf 333-334}, 13
  (2000), astro-ph/9909076; J.R.\ Mould \etal, \apj {\bf {529}}, 786
  (2000); W.L. Freedman \etal, astro-ph/0012376 (2000).
  
\bibitem{LSS} The LSS prior constrains the amplitude using cluster
  abundances, $\sigma_8 \Omega_m^{0.56}$=$0.55^{+.02,+.11}_{-.02,-.08}$,
  distributed as a Gaussian (first error) smeared by a uniform (tophat)
  distribution (second error). [See also, \eg J.P.\ Henry, \apj {\bf 534}
  565 (2000) which breaks this degeneracy somewhat.]  We constrain the
  shape of the power spectrum via $\Gamma + (n_s -
  1)/2$=$0.22^{+.07,+.08}_{-.04,-.07}$, where $\Gamma \approx \Omega_{m}
  \, h \, \, e^{-(\Omega_B(1+\Omega_{m}^{-1}(2{ h})^{1/2}) -0.06)}$.
  More details are in~\protect{\cite{BOOM2}} and J.R.\ Bond \& A.H.\ 
  Jaffe, Phil.\ Trans.\ R.\ Soc.\ London {\bf 357}, 57 (1999),
  astro-ph/9809043.
  
\bibitem{SNIa} A.G.\ Riess \etal, Astron.\ J.\ {\bf 116}, 1009 (1998); S.\ 
  Perlmutter \etal, \apj {\bf {517}}, 565 (1999).

\bibitem{strings} F.R.\ Bouchet, P.\ Peter, A.\ Riazuelo, M.\ 
  Sakellariadou, \prl submitted (2000), astro-ph/0005022. C.\ Contaldi,
  \prl submitted, astro-ph/0005115.  

\end{thebibliography}
\end{document}